Spectral Risk Measures and the Choice of Risk Aversion Function

By

Kevin Dowd and John Cotter[*]

Abstract

Spectral risk measures are attractive risk measures as they allow the user to obtain risk measures that reflect their risk-aversion functions. To date there has been very little guidance on the choice of risk-aversion functions underlying spectral risk measures. This paper addresses this issue by examining two popular risk aversion functions, based on exponential and power utility functions respectively. We find that the former yields spectral risk measures with nice intuitive properties, but the latter yields spectral risk measures that can have perverse properties. More work therefore needs to be done before we can be sure that arbitrary but respectable utility functions will always yield 'well-behaved' spectral risk measures.

March 11, 2007

[*] Kevin Dowd is at the Centre for Risk and Insurance Studies, Nottingham University Business School, Jubilee Campus, Nottingham NG8 1BB, UK; email: Kevin.Dowd@nottingham.ac.uk. John Cotter is at the Centre for Financial Markets, School of Business, University College Dublin, Carysfort Avenue, Blackrock, Co. Dublin, Ireland; email: john.cotter@ucd.ie. The authors would like to thank Carlo Acerbi and Dirk Tasche for fruitful conversations on the subject. Dowd's contribution was supported by an Economic and Social Research Council research fellowship on 'Risk measurement in financial institutions', and he thanks the ESRC for their financial support. Cotter's contribution to the study has been supported by a University College Dublin School of Business research grant.

## 1. Introduction

One of the most interesting and potentially most promising recent developments in the financial risk area has been the theory of spectral financial risk measures, recently proposed by Acerbi (2002, 2004). Spectral risk measures (SRMs) are closely related to the coherent risk measures proposed a little earlier by Artzner *et alia* (1997, 1999), and share with the coherent risk measures the highly desirable property of subadditivity (i.e., that the risk of the sum is no more than the sum of the individual risks[1]). It is also well-known by now that the most widely used risk measure, the Value-at-risk (VaR), is not subadditive, and the work by Artnzer *et alia* and Acerbi has shown that many (if not most) of the inadequacies of VaR as a risk measure can be traced to its non-subadditivity.

One of the nice features of SRMs is that they relate the risk measure itself to the user's risk-aversion – in effect, the spectral risk measure is a weighted average of the quantiles of a loss distribution, the weights of which depend on the user's risk-aversion function. SRMs can be applied to many different problems. For example, Acerbi (2004) suggests that they can be used to set capital requirements or obtain optimal risk-expected return tradeoffs, and Cotter and Dowd (2006) suggest that SRMs could be used by futures clearinghouses to set margin requirements that reflect their corporate risk aversion.

Spectral risk measures therefore enable us to link the risk measure to the user's attitude towards risk, the underlying objective being to ensure that if a user is more risk averse, other things being equal, then that user should face a higher risk, as given by the value of the SRM. However there is very little guidance on what a suitable risk aversion function might entail. For example, Szegö (2002) describes the process of multiplying coherent risk measures by an admissible risk aversion function but does not specify what an admissable risk aversion function might be. Moreover Acerbi (2004, p. 175) calls for the identification of suitable additional criteria over and above the coherent properties to assist the risk manager in choosing their risk aversion function, but he himself only illustrates one particular risk-aversion function – namely, an exponential one.

---

[1] More formally, if $\rho(.)$ is a measure of risk, and $A$ and $B$ are two positions, subadditivity means that $\rho(A+B) \leq \rho(A) + \rho(B)$. Subadditivity is a crucial condition because it ensures that our risks do not increase overall when we put them together. As Acerbi and others have pointed out, any risk measure that does not satisfy subadditivity has no real claim to be regarded as a 'respectable' risk measure at all (see, e.g., Acerbi (2004, p. 150)).

This paper investigates SRMs further. In particular, it examines two alternative types of SRM based on alternative underlying utility functions: an exponential SRM based on an exponential utility function, which is equivalent to the one that Acerbi studies, and a power SRM based on a power utility function. As far as we are aware, these latter SRMs have not been studied before. They are however a natural object of study as the power utility function is widely used. We find that the exponential utility function leads to 'well-behaved' risk-aversion functions and a 'well-behaved' SRM, but the power utility function does not. Indeed, we find that the power utility function can lead to a situation where an increase in risk aversion leads to a *decrease* in the value of the SRM – a clear sign of a 'badly behaved' SRM. These results are surprising, and suggest that we cannot simply pull a respectable utility function off the shelf, obtain its risk-aversion function and thence its SRM, and necessarily expect that this SRM will be 'well-behaved'.

The article is organised as follows. Section 2 sets out the essence of Acerbi's theory of spectral risk measures. Section 3 explains spectral risk measures based on exponential utility functions, and section 4 examines spectral risk measures based on a power utility function. Section 5 concludes.

**2. Spectral Risk Measures**

Following Acerbi (2004), consider a risk measure $M_\phi$ defined by:

(1) $$M_\phi = \int_0^1 \phi(p) q_p \, dp$$

where $q_p$ is the $p$ loss quantile and $\phi(p)$ is a user-defined weighting aversion function with weights defined over $p$, where $p$ is a range of cumulative probabilities $p \in [0,1]$. We can think of $M_\phi$ as the class of quantile-based risk measures, where each individual risk measure is defined by its own particular weighting function.

Two well-known members of this class are the VaR and the Expected Shortfall (ES). The VaR at the $\alpha$ confidence level is:

$$(2) \qquad VaR_\alpha = q_\alpha$$

The VaR places all its weight on a single quantile that corresponds to a chosen confidence level, and places no weight on any others, i.e., with the VaR risk measure, $\phi(p)$ takes the degenerate form of a Dirac delta function that gives the outcome $p = \alpha$ an infinite weight and gives every other outcome a zero weight.

For its part, the ES at the confidence level $\alpha$ is the average of the worst $1-\alpha$ of losses and (in the case of a continuous loss distribution) is:

$$(3) \qquad ES_\alpha = \frac{1}{1-\alpha} \int_\alpha^1 q_p \, dp$$

With the ES, $\phi(p)$ takes gives tail quantiles a fixed weight of $\frac{1}{1-\alpha}$ and gives non-tail quantiles a weight of zero.

A drawback with both of these risk measures is that they inconsistent with risk aversion in the traditional sense. This can be illustrated in the context of the theory of lower partial moments (see Bawa (1975), Fishburn (1977) and Grootveld and Hallerbach (2004)). Given a set of returns $r$ and a target return $r^*$, the lower partial moment of order $k \geq 0$ around $r^*$ is equal to $E\{[\max(0, r^* - r]^k\}$. The parameter $k$ reflects the user's degree of risk aversion, and the user is risk-averse if $k > 1$, risk-neutral if $k = 1$ and risk-loving if $0 < k < 1$. It can then be shown that the VaR is a preferred risk measure only if $k = 0$, i.e., the VaR is our preferred risk measure only if we are very risk-loving! The ES would be our preferred risk measure if $k = 1$, and this tells us that the ES is our preferred risk measure only if the user is risk-neutral between better and worse tail outcomes.

A user who is risk averse might prefer to work with a risk measure that take account of his/her risk aversion, and this takes us to the class of spectral risk measures (SRMs). In loose terms, an SRM is a quantile-based risk measure that takes the form of (1) where $\phi(p)$ reflects the user's degree of risk aversion. More precisely, we can consider SRMs as the subset of $M_\phi$ that satisfy the following properties of positivity, normalisation and increasingness due originally to Acerbi:[2]

1. *Positivity*: $\phi(p) \geq 0$.

2. *Normalisation*: $\int_0^1 \phi(p) dp = 1$.

3. *Increasingness*: $\phi'(p) \geq 0$.

The first coherent condition requires that the weights are weakly positive and the second requires that the probability-weighted weights should sum to 1, but the key condition is the third one. This condition is a direct reflection of risk-aversion, and requires that the weights attached to higher losses should be no less than the weights attached to lower losses. Typically, we would also expect the weight $\phi(p)$ to rise with $p$.[3] In a 'well-defined' case, we would expect the weights to rise smoothly, and the more risk-averse the user, the more rapidly we would expect the weights to rise.

A risk measure that satisfies these properties is attractive not only because it takes account of user risk-aversion, but also because such a risk measure is known to be coherent.[4]

There still remains the question of how to specify $\phi(p)$, and perhaps the most natural way to obtain $\phi(p)$ is from the user's utility function[5].

---

[2] See Acerbi (2002, 2004). However, it is worth pointing out that he deals with a distribution in which profit outcomes have a positive sign, whereas we deal with a distribution in which loss outcomes have a positive sign. His first condition is therefore a negativity condition, whereas ours is a positivity condition, but this difference is only superficial and there is no substantial difference between his conditions and ours.

[3] The conditions set out allow for the degenerate limiting case where the weights are flat for all p values, and such a situation implies risk-neutrality and is therefore inconsistent with risk-aversion. However, we shall rule out this limiting case by imposing the additional (and in the circumstances very reasonable) condition that $\phi(p)$ must rise over at least some point as $p$ increases from 0 to 1.

[4] This follows from Acerbi (2004, Proposition 3.4).

## 3. Exponential Spectral Risk Measures

Suppose, for example, that we specify the following exponential utility function defined over random outcomes $x$:

$$U(x) = -e^{-ax} \tag{4}$$

where $a > 0$ is the Arrow-Pratt coefficient of *absolute* risk aversion (ARA). The coefficients of absolute and relative risk aversion are:

$$R_A(x) = -\frac{U''(x)}{U'(x)} = a \tag{5a}$$

$$R_R(x) = -\frac{xU''(x)}{U'(x)} = xa \tag{5b}$$

We now set

$$\phi(p) = \lambda e^{-a(1-p)} \tag{6}$$

where $\lambda$ is an unknown positive constant. This clearly satisfies properties 1 and 3, and we can easily show (by integrating $\phi(p)$ from 0 to 1, setting the integral to 1 and solving for $\lambda$) that it satisfies 2 if we set

$$\lambda = \frac{a}{1 - e^{-a}} \tag{7}$$

---

[5] See also Bersimas *et alia* (2004).

Hence, substituting (7) into (6) gives us the exponential weighting function (or risk-aversion function) corresponding to (4):[6]

$$(8) \quad \phi(p) = \frac{ae^{-a(1-p)}}{1-e^{-a}}$$

This risk-aversion function is illustrated in Figure 1 for two alternative values of the ARA coefficient, $a$. Observe that this weighting function has a nice shape: for the higher $p$ values associated with higher losses, we get bigger weights for greater degrees of risk-aversion. In addition, as $p$ rises, the rate of increase of $\phi(p)$ rises with the degree of risk-aversion.

**Insert Figure 1 here**

The SRM based on this risk-aversion function, the exponential SRM, is then found by substituting (8) into (1), viz.:[7]

$$(9) \quad M_\phi = \int_0^1 \phi(p) q_p \, dp = M_\phi = \frac{a}{1-e^{-a}} \int_0^1 e^{-a(1-p)} q_p \, dp$$

We also find that the risk measure itself rises with the degree of risk-aversion, and some illustrative results are given in Table 1. For example, if losses are distributed as standard normal and we set $a = 5$, then the spectral risk measure is 1.0816. But if we increase $a$ to 25, the measure rises to 1.9549: the greater the risk-aversion, the higher the exponential spectral risk measure.

---

[6] Strictly speaking, Acerbi's proposition 3.19 in Acerbi (2004, p. 182) defines his weighting function in terms of a parameter $\gamma > 0$, but his weighting function and (7) are equivalent subject to the proviso that $\gamma = 1/a$.

[7] Estimates of (9) were obtained using Simpson's rule numerical quadrature with $n=10,000,001$ 'slices'. The actual calculations were carried out using the CompEcon function in MATLAB given in Miranda and Fackler (2002). When estimating spectral risk measures we find that estimates converge only slowly – and from below – as $n$ increases (see Dowd (2005, Table 3.4). It is therefore important to use a value of $n$ that produces estimates close to the limiting value of the SRM as $n$ becomes very large, and some trials with alternative values of $n$ suggested that $n=10,000,001$ is adequate for our purpose here. We also use $n=10,000,001$ rather than the more obvious $n=10,000,000$ because the Simpson's rule algorithm requires $n$ to be odd.

Insert Table 1 here

The relationship of the exponential SRM and the coefficient of absolute risk aversion is illustrated further in Figure 2. We can see that the risk measure rises smoothly as the coefficient of risk aversion increases, as we would expect.

Insert Figure 2 here

**4. Power Spectral Risk Measures**

We can also obtain SRMs based on other utility functions, and a popular alternative to the exponential utility function is the power utility function[8]:

$$(10) \qquad U(x) = \frac{x^{1-c}}{1-c}$$

where $0 < c < 1$ is the Arrow-Pratt coefficient of *relative* risk aversion (RRA). This function has a constant coefficient of relative risk aversion and so belongs to the family of Constant Relative Risk Aversion (CRRA) utility functions. Its coefficients of absolute and relative risk aversion are:

$$(11a) \qquad R_A(x) = -\frac{U''(x)}{U'(x)} = \frac{c}{x}$$

$$(11b) \qquad R_R(x) = -\frac{xU''(x)}{U'(x)} = c$$

We convert the utility function (10) into a risk-aversion function $\phi(p)$ using

$$(12) \qquad \phi(p) = \lambda \frac{(1-p)^{c-1}}{1-c}$$

---

[8] This power function is heavily used and represents a special case of the Hyberbolic Absolute Risk Aversion (HARA) family.

where $\lambda$ is another unknown constant. In this case, $\lambda > 0$ suffices to ensure that conditions 1 and 3 hold, and this risk-aversion function satisfies property 2 if we set

(13) $$\lambda = c(1-c)$$

Hence, the power utility function leads to the power risk-aversion function:

(14) $$\phi(p) = c(1-p)^{c-1}$$

The power risk-aversion function (13) is plotted in Figure 3 for illustrative $c$ values equal to 0.7 and 0.9. This shows that the $\phi(p)$ curve for the higher degree of RRA is initially higher than the $\phi(p)$ curve for the lower degree of RRA, but then falls below it. This tells us that with *higher* risk aversion, relatively *more* weight is placed on the *lower* losses and relatively less weight is placed on the *higher* losses, compared to the case with lower risk aversion! This is clearly odd, even though the $\phi(p)$ function satisfies the properties 1-3 above.

**Insert Figure 3 here**

The resulting risk measure (obtained by substituting (14) into (1)) is then

(15) $$M_\phi = \int_0^1 q_p \phi(p) dp = M_\phi = \int_0^1 q_p c(1-p)^{c-1} dp$$

This risk measure also satisfies conditions 1 to 3 above and so qualifies as an SRM as we have defined it. This also means that it is coherent.

Figure 4 shows a plot of the power SRMs against $c$ and Table 2 gives some illustrative numerical values for losses distributed as standard normal. The SRM plot is surprising: it starts from a value of 0 when $c$ is very close to 0, then rises sharply to peak at a $c$ value of about 0.11, and thereafter falls smoothly to 0, which it approaches as *c* approaches 1: the greater the risk-aversion, the lower the power spectral risk measure. So, for example, if losses are distributed as standard normal and we set $c = 0.1$, the spectral risk measure is 1.9278, but if we increase *c* to 0.9, the measure falls to 0.0968. The initial rise in SRM is exactly what we would expect of a 'well-behaved' SRM, but the subsequent fall from its peak value is surprising: the fall from its peak value tells us, paradoxically, that the SRM *falls* as the user becomes *more* risk-averse. The explanation for this curious result can be seen in Figure 3: the higher the coefficient of relative risk aversion, the more weight is placed on the low loss outcomes on the left-hand side and the lower the weight placed on the high loss outcomes on the right-hand side.

**Insert Figure 4 here**
**Insert Table 2 here**

Thus, we have a spectral risk measure that satisfies Acerbi's conditions, and yet the weighting function and resulting risk measure are manifestly 'badly-behaved' except for rather low values of $c$.[9] Conditions 1 to 3 are clearly not sufficient to ensure that we get a 'well-behaved' risk aversion function or a 'well-behaved' SRM.

These findings suggest that we might wish to change the conditions required of an SRM. For example, we might impose additional 'well-behavedness' conditions on the risk-aversion function. However, the cost of doing so would be to restrict our choice of risk-aversion function, i.e., so we could not pull any risk-aversion function that satisfies 1 to 3 off the shelf and input it into (1) to get a 'well-behaved' risk measure. Alternatively, we might still wish to use a power SRM if we were confident that we had a *c* value that lay within the range where the power SRM is 'well-behaved'.

However, neither of these suggestions resolves the underlying problem: if we have a power utility function with an arbitrary coefficient of relative risk aversion in the maximum permissible range of (0,1), then how should we obtain a 'well-behaved' SRM that reflects this utility function's aversion to risk?

---

[9] This example based on a power function also illustrates that at least some HARA functions will also produce 'badly-behaved' risk measures.

## 5. Conclusions

This paper has examined spectral risk measures based on exponential and power utility functions. We find that the exponential utility function leads to risk-aversion functions and spectral risk measures with intuitive and nicely behaved properties. However, we find that the same is not true of the power utility function. This produces risk-aversion functions and related risk measures that are patently unsatisfactory. Indeed, for some ranges of the coefficient of relative risk aversion we find that an increase in risk aversion leads to a decrease in the value of the power SRM. This result is consistent with the definition of an SRM – and with the axioms of coherence too – but runs very much counter to the spirit of spectral risk measures, which is that SRMs should reflect risk aversion in a non-perverse way.

These power utility findings also indicate that we cannot assume any risk aversion function that we like, and necessarily get well-behaved risk measures: after all, the power function is quite a reasonable one, and yet it produces very unreasonable spectral risk measures. Thus, the choice of risk aversion functions needs to be restricted if we are to be sure of getting well-behaved spectral risk measures from arbitrary but otherwise respectable utility or risk-aversion functions. How best to do that is a good subject for future research.

**FIGURES**

**Figure 1: Exponential Risk Aversion Functions**

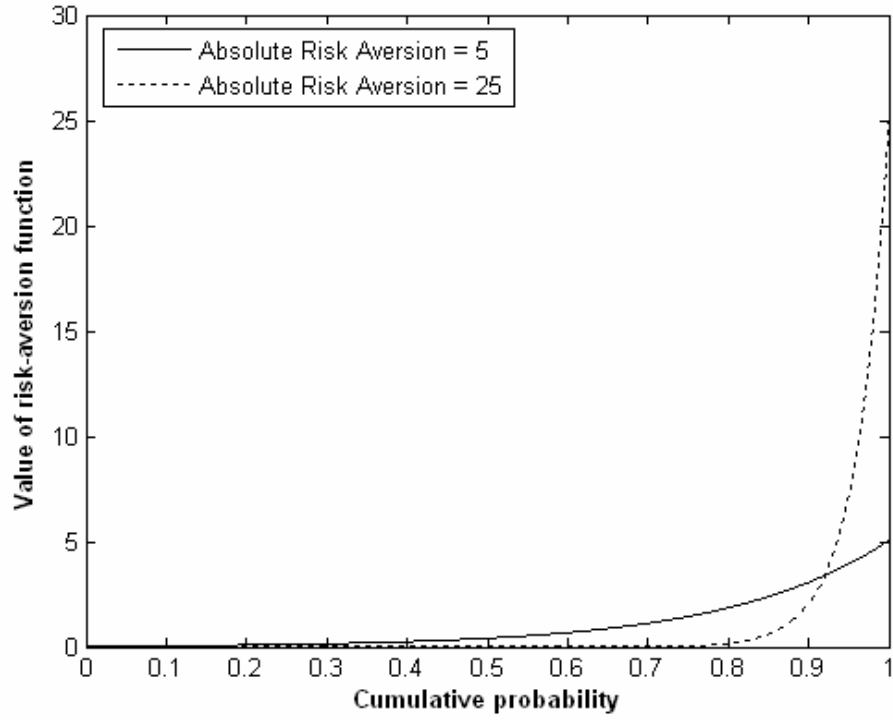

Notes: Weights are based on the exponential risk-aversion function (8).

**Figure 2: Plot of Exponential Spectral Risk Measure Against the Coefficient of Absolute Risk Aversion: Standard Normal Loss Distribution**

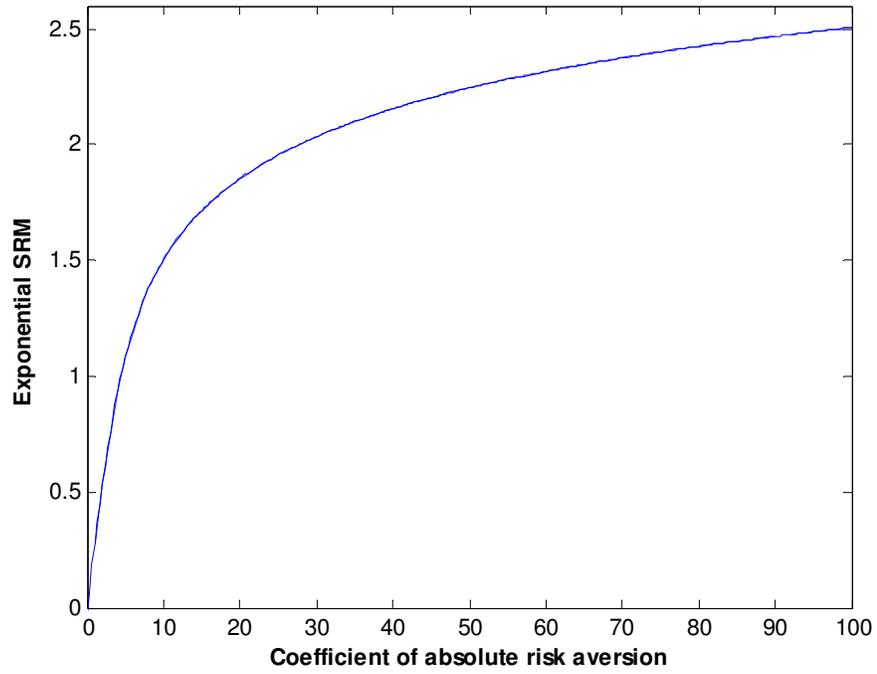

Notes: As per Notes to Table 1.

**Figure 3: Power Risk Aversion Functions**

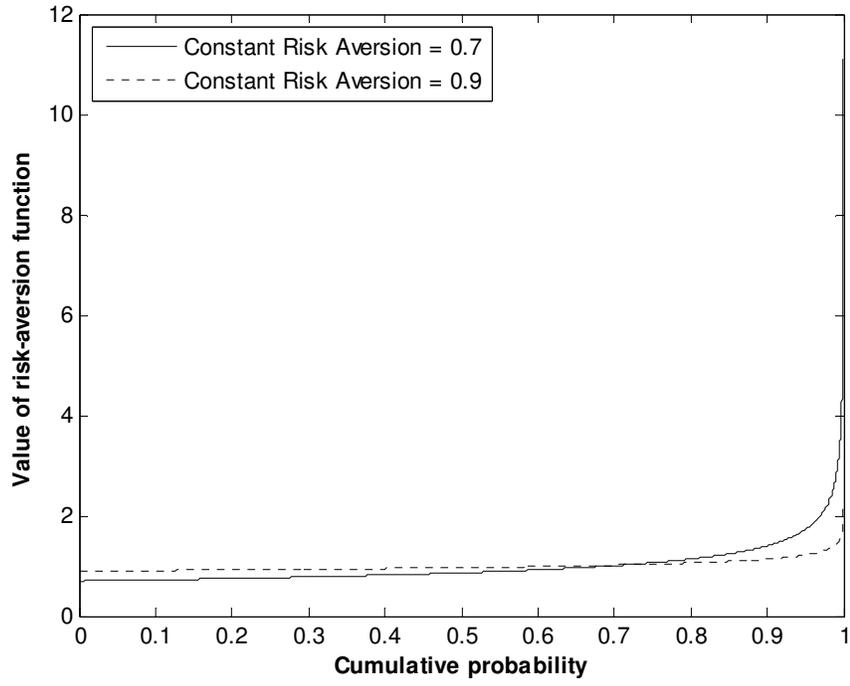

Notes: Weights are based on the power risk-aversion function (14).

**Figure 4: Plot of Power Spectral Risk Measure Against the Coefficient of Relative Risk Aversion: Standard Normal Loss Distribution**

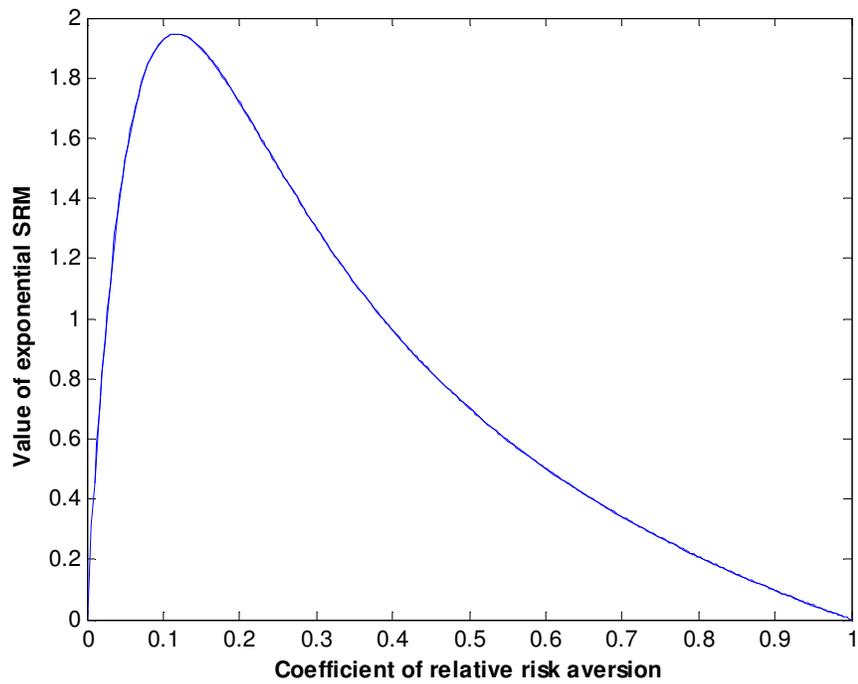

Notes: As per Notes to Table 2.

**Table 1: Values of Exponential Spectral Risk Measure with Standard Normal Losses**

| Coefficient of Absolute Risk Aversion | Exponential Spectral Risk Measure |
|---|---|
| 1 | 0.2781 |
| 5 | 1.0816 |
| 25 | 1.9549 |
| 100 | 2.5055 |

Notes: Estimates are of (9) obtained using Simpson's rule numerical quadrature with $n$=10,000,001 'slices'. The actual calculations were carried out using the CompEcon function in MATLAB given in Miranda and Fackler (2002).

**Table 2: Values of Exponential Spectral Risk Measure with Standard Normal Losses**

| Coefficient of Relative Risk Aversion | Power Spectral Risk Measure |
|---|---|
| $\to 0$ | 0 |
| 0.1 | 1.9278 |
| 0.5 | 0.7026 |
| 0.9 | 0.0968 |
| $\to 1$ | 0 |

Note: Estimates are of (15) obtained using Simpson's rule numerical quadrature with $n$=10,000,001 'slices'. The actual calculations were carried out using the CompEcon function in MATLAB given in Miranda and Fackler (2002).